\documentclass[aps,pra,superscriptaddress,twocolumn, 10pt]{revtex4-1}
\pdfoutput=1

\usepackage[utf8]{inputenc}
\usepackage{amssymb,amsmath}
\usepackage{graphicx}
\usepackage{wrapfig}
\usepackage[usenames,dvipsnames]{xcolor}
\usepackage[colorlinks,bookmarks=false,citecolor=blue,linkcolor=black,urlcolor=blue]{hyperref}
\usepackage[vcentermath]{youngtab} 
\usepackage{young}
\usepackage{capt-of}

\newcommand{\ignore}[1]{} 
\newcommand{\SU}{\ensuremath{\mathrm{SU}}}

\newcommand{\U}{\ensuremath{\mathrm{U}}}
\newcommand{\A}{\mathrm{A}}
\newcommand{\B}{\mathrm{B}}

\newcommand{\CC}{\mathbb{C}}
\newcommand{\HH}{\mathcal{H}}
\newcommand{\kk}{\ensuremath{\mathcal{K}}}

\begin{document}

\title{Quantum phases of collective SU(3) spin systems with bipartite symmetry}

\author{D\'avid Jakab}
\affiliation{Wigner Research Centre for Physics,  H-1525 Budapest P.O. Box 49, Hungary}
\affiliation{Institute of Physics - University of Pécs, H-7624 Pécs, Ifjúság u.~6, Hungary}

\author{Zolt\'an Zimbor\'as}
\affiliation{Wigner Research Centre for Physics,  H-1525 Budapest P.O. Box 49, Hungary}
\affiliation{BME-MTA Lend\"ulet Quantum Information Theory Research Group, Budapest, Hungary}
\affiliation{Mathematical Institute, Budapest University of Technology and Economics, \\ H-1111 Budapest P.O.Box 91, Hungary}
\date{\today}

\begin{abstract}
We study a bipartite collective spin-$1$ model with exchange interaction between the spins. The bipartite nature of the model manifests itself by the spins being divided into two equal-sized subsystems; within each subsystem the spin-spin interactions are of the same strength, across the subsystems they are also equal, but the two coupling values within and across the subsystem are different. Such a set-up is inspired by recent  experiments with ultracold atoms. Using the $\SU(3)$-symmetry of the exchange interaction and the permutation symmetry within the subsystems, we can employ representation theoretic methods to diagonalize the Hamiltonian of the system in the entire parameter space  of the two coupling-strengths. These techniques then allow us to explicitly construct and explore the ground-state phase diagram. The phase diagram turns out to be rich containing both gapped and gapless phases. An interesting observation is that one of the five phases features a strong bipartite symmetry breaking, meaning that the two subsystems  in the ground states are in different $\SU(3)$ representations.
\end{abstract}

\maketitle

\section{Introduction}

Since the beginning of quantum many-body physics, spin lattice system with rotational invariant, i.e., $\SU(2)$ symmetric, interaction terms have received a special attention.  From a theoretical point of view, it was a natural generalization to also consider quantum spin models where $\SU(2)$ is enlarged to the symmetry group $\SU(N)$ with ${N >2}$ \cite{sutherland1975model,read1989valence,marston1989large}. A particular application was the case of materials described by spin models with orbital degeneracy yielding an $\SU(4)$ symmetric point \cite{kugel1973crystal, arovas1995tetrahis, pati1998alternating, li1998su(4), penc2003quantum, tokura2000orbital}, but the main motivation to study $\SU(N)$ spin systems remained mainly formal. In particular, one of the driving forces behind the theoretical studies was the realization that $\SU(N)$ symmetric spin models have very rich phase diagrams \cite{read1989valence, marston1989large, harada2003neel, assaad2005phase, greiter2007valence, rapp2008trionic, xu2008resonating}. Later these studies gained an unexpected experimental relevance with the advent of experiments with ultracold atomic systems. For example, in the ground state and certain excited states of alkaline-earth atoms, the nuclear spin $I$ is almost perfectly decoupled from the electronic angular momentum, and in an optical lattice the interaction between the trapped atoms is governed only by the electronic structure. Therefore, in a system of trapped alkaline-earth atoms the s-wave scattering lengths are independent of the total nuclear spin of the colliding atoms \footnote{The nuclear spin still plays a part in the scattering indirectly via the fermionic statistics.}, which results in the $SU(2I+1)$ symmetry of the effective Hamiltonian describing the interaction \cite{gorshkov2010two-orbital}.
  
Another consequential effect of ultracold atom experiments on many-body physics is an enhanced focus on long-range systems, since in some of these experiments the interactions  decay as an inverse power-law \cite{porras2004effective} or are even of mean-field type \cite{beverland2016realizing}. This opens the way to experimentally realize long-range models and scenarios that were previously only thought of being of theoretical interests, e.g., Curie-Weiss-type transverse-field Ising models (i.e., the Lipkin-Meshkov-Glick model) \cite{zibold2010classical,gang2009interaction,morrison2008dynamical,cirac1998quantum} or the Haldane-Shastry model \cite{hung2016quantum, grass2014trapped}; and to study phenomena that are not possible in short-range models, e.g., breaking of continuous symmetries in one-dimensional systems.

In this paper, we investigate an interplay of high internal symmetries and the long-ranged property for a collective system of $\SU(3)$ spins that are divided in a bipartite way into two subsystems. The $\SU(3)$-invariant interaction within the subsystems are of Curie-Weiss-type, i.e., each spin interacts with each of the other spins with the same strength. The spin-spin interactions across the subsystems are also equal, however, their strength is different than of those within the subsystems. This way, we introduced a bipartite structure in a mean-field type model, which should make the antiferromagnetic region more interesting and the phase structure richer. Such a system may look quite artificial for the first glance, however, experimental techniques with ultracold atoms and cavity electrodynamics represent a promising way towards its realization. One may expect, that a dual system of ultracold ensembles inside lossy optical cavities \cite{Xu2014Synchronization, Roth2016Synchronization, Weiner2017Phase} can actually be used to realize \SU(3) symmetric Mott insulators on a bipartite lattice, where the permutation invariant infinite-range interaction is provided by the cavity photons. The two ensembles can be realized by different electric configurations of the ultracold atomic gas, e.g., the two 3-component $F=1$ hyperfine states on the two sides of the rubidium $D_1$ transition \footnote{D. A. Steck, ``Rubidium 87 D Line Data'', available online at \url{http://steck.us/alkalidata} (revision 2.2.1, 21 November 2019).}.

The paper is structured as follows: In Section~\ref{sec:model}, we define the model and introduce the notation used in the rest of the work. This is followed by a formal diagonalization of the Hamiltonian using representation theoretic tools in Section~\ref{sec:eigenspaces}. In Section~\ref{sec:phase_diag}, we explicitly construct and explore the ground-state phase diagram. Finally, we summarize our results and provide an outlook in Section~\ref{sec:summary}.

\section{The model Hamiltonian} \label{sec:model}
A large part of interacting spin systems are described by Hamiltonians which
factorize into two-particle contributions
\begin{equation}
  \label{eq:Ham}
  H=\sum_{\lbrace(i,j)\rbrace} H_{i j},
\end{equation}
where $\lbrace(i,j)\rbrace$ denotes the ``neighbors'' of the underlying lattice or
graph, i.e., those spins on sites $i$ and $j$ which interact with each other.
In the case of the spin-$\tfrac{1}{2}$ Heisenberg model, the
two-particle interaction $H_{i j}$ is described by the rotation invariant term,
\begin{equation}
  \label{eq:spin12hij}
  H_{ij}=2 J\, \mathbf{S}_i\cdot \mathbf{S}_j = J \mathcal{P}_{ij} + \text{const},
\end{equation}
where $\mathbf{S}_i$ is the vector of spin-$\tfrac{1}{2}$ operators. $H_{ij}$ is also called the exchange interaction, since it
can  be expressed by the swap operator $\mathcal{P}_{ij}$  that exchanges sites $i$ and $j$.

There are several ways to generalize the interaction defined by Eq.~\eqref{eq:spin12hij} for
higher spin systems. One may, for example, look for higher-dimensional
representations of the $\SU(2)$ spin algebra and keep imposing the rotation invariance. In particular, for a system with three spin components we obtain this way the spin-1 bilinear-biquadratic model \cite{papanicolaou1986ground, fath1991period}.
Another quite natural way of generalizing Eq.~\eqref{eq:spin12hij} is keeping the interaction's exchange
nature by considering two-particle Hamiltonians that are proportional to the swap operator.  This way the original global $\SU(2)$ symmetry of the model is extended to $SU(N)$.

In this paper we follow the latter route and consider the $\SU(3)$ symmetric exchange interaction. The Hilbert space corresponding to each site has three basis states which form the defining representation of $\SU(3)$, while the two-site Hilbert space decomposes into two irreducible subspaces under global $\SU(3)$ transformations. As a consequence, on two sites, the linear span of two independent $\SU(3)$ invariant operators encompasses all operators of such nature. The two invariant operators we choose are the identity and a quadratic expression of the conventional two-site $\SU(3)$ generators, the quadratic Casimir operator
\begin{equation}
  \label{eq:twositecasimir}
  C_{ij}=\sum_{\alpha,\beta=1}^3(S_i^{\alpha\beta}+S_j^{\alpha\beta})(S_i^{\beta\alpha}+S_j^{\beta\alpha}).
\end{equation}
The generators $S_i^{\alpha\beta}$ ($\alpha,\beta=1,2,3$) act on the local basis states of site $i$ 
as matrix units that are made traceless
\begin{equation}
  S^{\alpha\beta}_i=|\beta\rangle\langle \alpha |-\frac{1}{3}\text{Tr}\left( |\beta\rangle\langle \alpha | \right).
\end{equation}
Out of the three diagonal ones only two are independent, since $\sum_{\alpha=1}^3 S_i^{\alpha\alpha}=0$. 
The non-diagonal generators are not hermitian, instead we have $(S_i^{\alpha\beta})^\dagger=S_i^{\beta\alpha}$. Such a pair acts as a raising and a lowering operator between the basis states $|\alpha\rangle$ and $|\beta\rangle$.  
The generators fulfill the $\SU(3)$ commutation relations
\begin{equation}
  \left[ S_i^{\alpha\beta}, S_i^{\gamma\delta}\right]=\delta_{\alpha\delta}S_i^{\gamma\beta}-\delta_{\beta\gamma}S_i^{\alpha\delta},
\end{equation}
using this we conclude that \eqref{eq:twositecasimir} indeed commutes with all the $\SU(3)$ generators.
Since the exchange operator is invariant with respect to global $\SU(3)$ transformations, the exchange interaction takes the form
\begin{equation}
  \label{eq:hijsu3}
  H_{ij}{=}J \mathcal{P}_{ij}{=} J \sum_{\alpha, \beta} S^{\alpha \beta}_i S^{\alpha \beta}_j {+} \text{const}{=}\frac{J}{2}C_{ij}{+} \text{const}'.
\end{equation}
In the following we drop the constant terms. The model Hamiltonian obtained from such a two-particle interaction 
can also be thought of as a special case of the spin-1 bilinear biquadratic model corresponding to its high-symmetry points \cite{penc2011spin}.

The simplest case in which our model is exactly diagonalizable is when every spin interacts with every other spin with the same strength, in other words, when the graph underlying the two-particle interactions in Eq.~\eqref{eq:Ham} is the complete graph. Under the same circumstances the spin-$\frac{1}{2}$ Heisenberg Hamiltonian of Eq.~\eqref{eq:spin12hij} reduces to the square of the total spin operator. Analogously, a Hamiltonian on a complete graph with two-body $\SU(3)$-symmetric terms of Eq.~\eqref{eq:hijsu3} will be proportional to the quadratic Casimir operator of the global $SU(3)$ spin operators, and hence its eigenproblem simplifies to determining how the entire Hilbert space decomposes into  $\SU(3)$ irreducible representations (irreps). This was discussed in \cite{hamermesh1989group} and also in \cite{jakab2018bilinear} as a special case of the bilinear-biquadratic model on the complete graph.

 In order to facilitate bipartite symmetry (and its possible violation),  we partition the complete graph into two subsystems denoted by A and B. 
The strength of the interaction between two arbitrary spins on the same subsystems is set to $J_1$, and on different subsystems to $J_2$.  The Hamiltonian describing the entire system reads as:
\begin{equation}
  \label{eq:Hambip}
  H=(J_1-J_2)\sum_{\substack{i,j\in \A \\ i< j}}C_{ij} +
  (J_1-J_2)\sum_{\substack{i,j\in \B \\ i< j}}C_{ij} +
  J_2\sum_{\substack{i,j\in \mathrm{AB} \\ i<j}}C_{ij}.
\end{equation}
We introduce the parameter $\theta$ with $\text{tan}(\theta)=J_2/(J_1-J_2)$ and rescale the Hamiltonian in order to measure the energy in units of $\sqrt{J_1^2+2J_2^2-2J_1J_2}$. We also introduce the quadratic Casimir operators on of the two subsystems $C_\A$, $C_\B$, and the entire Hilbert space $C_\mathrm{AB}$
\begin{equation}
C_X=\sum_{\alpha,\beta=1}^3\left(  \sum_{i\in X}S_i^{\alpha,\beta}\right)\left( \sum_{j\in X}S_j^{\beta\alpha} \right).
\end{equation}
With these the Hamiltonian \eqref{eq:Hambip} takes the form
\begin{equation}
  \label{eq:HBLBQB}
  H_{\mathrm{CBE}} = \sin(\theta) C_{\mathrm{AB}} + \cos(\theta) \left(C_{\mathrm{A}} + C_{\mathrm{B}}\right).
\end{equation}
It is assumed, that the subsystems $\A$ and $\B$ are identical, each having $N$ sites, and therefore the
bipartite symmetry in Eq.~\eqref{eq:HBLBQB} is explicit along with the $\SU(3)$ symmetry, and the spins act in a mean-field-like collective manner.
Thus, throughout the paper we will call it the spin-$1$ {\it collective bipartite exchange Hamiltonian}, or  {\it CBE Hamiltonian} for short.

\section{Eigenspaces of the Hamiltonian}
\label{sec:eigenspaces}
In this section, we introduce some necessary concepts and provide the decomposition of the Hilbert space into the eigenspaces of the CBE Hamiltonian.

\subsection{Eigenspace decomposition}

The Hilbert space $\HH_{\A\B}\cong (\CC^3)^{\otimes 2N}$ decomposes into a direct sum of irreducible subspaces under global $\SU(3)$ transformations. The Hilbert spaces $\HH_{\mathrm{A}}\cong\HH_{\mathrm{B}}\cong (\CC^3)^{\otimes N}$ also have a similar decomposition under their respective N-fold $\SU(3)$ transformations, or more explicitly:

\begin{multline}
  \label{eq:decomp}
\HH_{\mathrm{AB}}\cong\HH_{\A}\otimes\HH_{\B}\cong
\bigoplus_{\mu\in 2\mathrm{N}}\kk^{(\mu)}_{\A\B}\otimes \HH^{(\mu)}_{\A\B}\cong\\
\left( \bigoplus_{ \mu\in \mathrm{N} }\kk^{(\mu)}_{\A}\otimes \HH^{(\mu)}_{\A} \right)\otimes\left( \bigoplus_{\mu\in \mathrm{N}}\kk^{(\mu)}_{\B}\otimes \HH^{(\mu)}_{\B} \right).
\end{multline}
Here $\HH^{(\mu)}$ are subspaces where the respective N-fold or 2N-fold $\SU(3)$ transformations act irreducibly, and $\kk^{(\mu)}$ are subspaces where the same transformations act as identity. The dimensions of these $\kk^{(\mu)}$ subspaces are equal to the multiplicities of the $\SU(3)$ irreducible representation $\mu$ in the irrep decomposition of the 2N-fold (AB) or N-fold (A, B) direct product of the defining representation. Likewise, the $\mu$ irreps that appear in the direct sums are the same irreps appearing in these decompositions.

The eigenspaces of  $C_{\mathrm{AB}}$, $C_{\mathrm{A}}$ and $C_{\mathrm{B}}$ are precisely these subspaces of the form $\kk^{(\mu)}\otimes \HH^{(\mu)}$ in the decomposition of corresponding Hilbert spaces, thus, the diagonalization of the CBE Hamiltonian \eqref{eq:HBLBQB} turns into a representation theoretic problem. Since the Casimir operators appearing in the CBE Hamiltonian commute with each other, their eigenspaces must be compatible. 
This compatibility manifests by the direct products of $C_\A$ and $C_\B$ eigenspaces decomposing into direct sums of $C_{\A\B}$ eigenspaces in the following way.

\begin{multline}
\left(  \kk^{(\mu_\A)}_{\A}\otimes \HH^{(\mu_\A)}_{\A}\right) \otimes\left(  \kk^{(\mu_\B)}_{\B}\otimes \HH^{(\mu_\B)}_{\B}\right)\cong\\ \kk^{(\mu_\A)}_\A\otimes\kk^{(\mu_\B)}_\B\otimes\left(  \bigoplus_{\mu_{\A\B}}\HH^{(\mu_{\A\B})}_{\A\B}\right).
 \end{multline}
 As a result, we are able to label the eigenspaces of the CBE Hamiltonian  by $(\mu_\A, \mu_\B, \mu_{\A\B})$ triplets of $\SU(3)$ irreps. In this sense, however, not all $\SU(3)$ irreps are compatible with each other; the properties by which the valid triplets can be identified are:
 \begin{enumerate}
 \item $\mu_\A$ and $\mu_\B$ must appear in the irrep decomposition of the N-fold direct product of the defining representation of $\SU(3)$, $\mu_{\A\B}$ must appear in the decomposition of the 2N-fold direct product. 
 \item $\mu_{\A\B}$ must appear in the irrep decomposition of the direct product of $\SU(3)$ irreps $\mu_\A$ and $\mu_\B$.
 \end{enumerate}
 
 In order to be able to tell whether a particular triplet of $\SU(3)$ irreps has these properties, we need to state the exact rules for how a direct product of two arbitrary irreps decomposes into a direct sum of irreps. The irreps of $\SU(3)$ are traditionally labeled by two-row Young diagrams and when it is appropriate we will refer to irreps as diagrams, however, generally we will use another, equivalent labelling, the so-called Dynkin labels. This labelling consist of pairs of non-negative integers $\mu=(p, q)$, where $q$ is the length of the second row of the corresponding Young diagram, and $p$ is the difference between the lengths of the first row and the second row of the diagram \cite{iachello}, for example
 \begin{equation}
   \yng(3,3)\equiv (0,3)\qquad \yng(3,1)\equiv (2,1).
 \end{equation}
 The most widespread method to obtain the irreducible decomposition of the direct product of two such irreps is provided by the combinatorial Littlewood-Richardson rules \cite{sagan2013symmetric}; however, it is possible to prove that this irrep decomposition is equivalent to the following closed formula \cite{schlosser1987closed}:
 
 \begin{equation}
   \label{eq:productdecomposition}
   \begin{gathered}
     \begin{aligned}
       &(p_\A, q_\A)\otimes (p_\B, q_\B)\cong\\
       &\bigoplus_{i=0}^{i_1}\bigoplus_{k=0}^{k_1}\bigoplus_{l=l_0}^{l_1}(p_\A{+}p_\B{-}i-2k{+}l, q_\A{+}q_\B-i{+}k{-}2l),
     \end{aligned}\\
     \hspace*{-3mm} \; \, i_1{=}\text{min}\{p_\B,q_\A\},\quad k_1{=}\text{min}\{p_\A, p_\B{+}q_\B{-}i\},\\ \; \; \;  l_1{=}\text{min}\{q_\A{+}k{-}i,q_\B\},\quad l_0{=}\text{max}\{0,k{+}i{-}p_\B\}.
   \end{gathered}
 \end{equation}
 
 Using this formula, we can derive which irreps appear in the N-fold tensor product of the defining representation. For this, consider the tensor product of an arbitrary irrep $(p, q)$ with the defining representation; in this case Eq.~\eqref{eq:productdecomposition} reduces to
 \begin{align}
   & (p, q)\otimes (1, 0)\cong \nonumber \\ 
   & (p{+}1, q)\oplus (1{-}\delta_{p,0})(p{-}1, q{+}1)\oplus (1{-}\delta_{q,0})(p, q{-}1), \label{eq:fundamentalrepproduct}
 \end{align}
 where the numbers before the irreps represent their multiplicities in the decomposition. By iterating this step $(\text{N}-1)$ times, starting from $(p, q)=(1, 0)$ we arrive at  \begin{equation}
   \label{eq:hspacedecomposition}
   (1, 0)^{\otimes N}\cong \bigoplus_{v = 0}^{\lfloor N/3\rfloor}\bigoplus_{\substack{p,q \\ 2q+p=N-3v}}m_{pq}  \, \, (p, q),\qquad m_{pq}\ge 1 . 
 \end{equation}
 The exact values of the $m_{pq}$ multiplicities are harder to derive and not needed here, they can be calculated by using the hook length formula on the Young diagram corresponding to $(p, q)$ \cite{sagan2013symmetric}.
 
 \subsection{The ground-state subspace}
 
Now that we have described how to characterize the eigenspaces the CBE Hamiltonian, we move on to determine the triplet of \SU(3) irreps, $(\mu_\A, \mu_\B, \mu_{\A\B})$, that corresponds to the subspace of the ground states, i.e., the lowest energy eigenspace. We progress towards this goal through the following steps.  We fix two arbitrary irreps on the subsystems, $\mu_A$ and $\mu_B$, and then determine the irrep $\mu^{\text{opt}}_{\A\B}(\mu_\A, \mu_\B, \theta)$ which appears in the decomposition of $\mu_\A\otimes\mu_\B$ and minimizes  the term proportional to $C_{\A\B}$ in the CBE Hamiltonian \eqref{eq:HBLBQB}. Depending on the sign of the sine prefactor, this is equivalent to finding the irrep $\check{\mu}_{\A\B}(\mu_\A, \mu_\B)\cong(\check{p}_{\A\B}, \check{q}_{\A\B})$ that minimizes or the irrep $\hat{\mu}_{\A\B}(\mu_\A, \mu_\B)\cong(\hat{p}_{\A\B}, \hat{q}_{\A\B})$ that maximizes the eigenvalue of $C_{\A\B}$. After $\mu^{\text{opt}}_{\A\B}$ is known, the problem of determining the ground-state subspace reduces to finding irreps $\mu_\A$ and $\mu_\B$ for which the triplet  $(\mu_\A, \mu_\B, \mu_{\A\B}^{\text{opt}}(\mu_\A, \mu_\B, \theta))$ minimizes the eigenvalue of the CBE Hamiltonian.

 The eigenvalue of the quadratic Casimir operator of $\SU(3)$ corresponding to an arbitrary irrep $\mu=(p, q)$ is given by
 \begin{equation}
   c(p, q)=\frac{2}{3}(p^2+q^2+pq+3p+3q). 
   \label{eq:Casimir}
 \end{equation}
 In Appendix \ref{appendix:eigenvalues}, we derive the irreps, $\hat{\mu}_{\A\B}(\mu_\A, \mu_\B)$ and $\check{\mu}_{\A\B}(\mu_\A, \mu_\B)$ which maximize and minimize this eigenvalue. The Young diagram of $\hat{\mu}_{\A\B}(\mu_\A, \mu_\B)$ can be obtained by joining together the diagrams of $\mu_\A$ and $\mu_B$ row by row, 
 \begin{equation}
   \label{eq:maxirrep}
(\hat{p}_{\A\B},\hat{q}_{\A\B})=(p_\A+p_\B, q_\A+q_\B).
\end{equation}
The irrep $\check{\mu}_{\A\B}(\mu_\A, \mu_\B)$, however, can only be expressed using case distinction depending on $\mu_{\A}$ and $\mu_{\B}$. We introduce $X=p_\A-q_\B$ and $Y=p_\B-q_\A$. With these,
\begin{multline}
  \label{eq:cabmincases}
  (\check{p}_{\A\B}, \check{q}_{\A\B})=
  \begin{cases}
    (X-Y, Y)\quad\text{if}\  Y>0\ \text{and}\  X>Y \\
    (Y-X, X)\quad\text{if}\  X>0\ \text{and}\  X\le Y\\
    (Y, -X)\quad\text{if}\  X\le 0\ \text{and}\  Y>0\\
    (X, -Y)\quad\text{if}\  X>0\ \text{and}\  Y\le 0\\
    (-X, X-Y)\quad\text{if}\  X\le 0\ \text{and}\  X>Y\\
    (-Y, Y-X)\quad\text{if}\  Y\le 0\ \text{and}\  X\le Y.
  \end{cases}
\end{multline}

Out of the six potential $(\check{p}_{\A\B}, \check{q}_{\A\B})$ pairs, the ones in which both elements are non-negative are always identical.

\section{The ground-state phases} \label{sec:phase_diag}

In this section, using the results of Section~\ref{sec:eigenspaces}, we  determine the different ground-state phases of our model. It turns out that there are 5 distinct phases. The model becomes gapless in two phases and at the phase-boundaries, while it remains gapped within the other phases. Interestingly, in one of the gapped phases the bipartite sub-lattice symmetry is broken in a strong sense: 
namely, for the ground-state subspace the irreps $\mu_A$ and $\mu_B$ corresponding to the two subsystems are non-identical.

\begin{figure}[h!]
\includegraphics[width=0.24\textwidth]{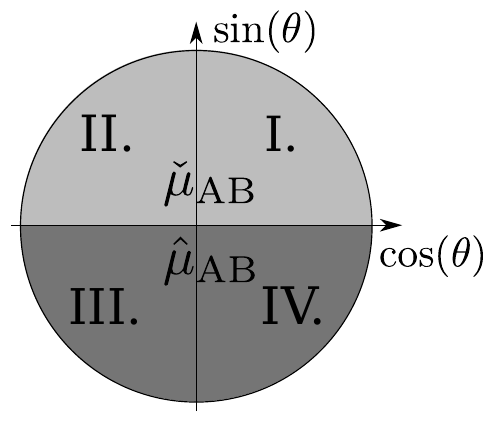}
\caption{The numbering of the quarters of the parameter region with the corresponding choice of the irrep $\mu_{\A\B}^{\text{opt}}$.}
\label{fig:quarterdiag}
\end{figure}

In  Section~\ref{sec:eigenspaces}, the CBE Hamiltonian was diagonalized and the optimal product irrep $\mu^{\text{opt}}_{\A\B}(\mu_A, \mu_B, \theta)$ was determined.  Thus, the identification of the ground-state subspace simplifies to finding the irreps $\mu_\A$ and $\mu_\B$ for which the triplet $(\mu_\A, \mu_\B, \mu_{\A\B}^{\text{opt}}(\mu_\A, \mu_\B, \theta))$ has minimal energy.
With $\mu_\A$ and $\mu_\B$ fixed, $\mu_{\A\B}^{\text{opt}}$ depends only on the sign of the prefactor of the Casimir operators $C_A$, $C_B$, and $C_{AB}$, i.e., on the sign  of $\sin(\theta)$ and $\cos(\theta)$. Hence,  it is instructive to investigate the ground state separately in the four quarters of the domain of our angle parameter $\theta$; we number these quarters clockwise, starting with $0\le\theta\le\pi/2$, as seen in Fig.~\ref{fig:quarterdiag}. 

Our task is to find for all values of $\theta$ the minimum of the energy as a function the four parameters $p_\A,q_\A,p_\B$ and $q_\B$.
It will be sometimes convenient to switch from these standard parameters of $\SU(3)$ irreps to a different set: 
\begin{equation}
  p=N v x,\quad q=\frac{N v}{2} (1-x).
  \label{eq:variabletransform}
\end{equation}
Here, $N v$ describes the number of boxes in the diagram; $v$ takes values between $0$ and $1$ in steps of $1/N$, and for every fixed value of $v$, $x$ takes values between $\text{mod}(Nv,2)/(Nv)$ and $1$ in steps of $2/Nv$. When taking the  thermodynamic limit, $N \to \infty$, these variables can be treated as continuous.
By substituting Eq.~\eqref{eq:variabletransform} into Eq.~\eqref{eq:Casimir}, one can see that the eigenvalue of the Casimir operator corresponding to $(p,q)$ has terms that are either quadratic or linear in $N$. In the thermodynamic limit it is sufficient to consider only the contributions of the quadratic part,
 \begin{equation}
   c_{qu}(p, q)=\frac{2}{3}(p^2+q^2+pq). 
 \label{eq:Casimirquadpart}
 \end{equation}
Coincidentally, for $(\check{p}_{\A\B}, \check{q}_{\A\B})$ this quadratic part of the Casimir is described by the same expression in all the cases of Eq.~\eqref{eq:cabmincases},
\begin{multline}
  \label{eq:cabmin}
  c_{qu}(\check{p}_{\A\B}, \check{q}_{\A\B})=\\c_{qu}(p_\A, q_\A)+c_{qu}(p_\B, q_\B)-(2p_\A q_\B+2p_\B q_\A+p_\A p_\B+q_\A q_\B).
\end{multline}

\subsection{First quarter $( 0<\theta<\pi/2)$} 
It is simplest to determine the ground state when the signs of the sine and cosine prefactors in the CBE Hamiltonian \eqref{eq:HBLBQB}, are the same, because in this case there is no competition between the interactions across and within the subsystems. Thus, we start with the region  $0<\theta<\pi/2$, i.e., the first quarter. In this region of the parameter space the eigenvalues of all the Casimir operators need to be minimized. From Eq.~\eqref{eq:Casimir} one can immediately see that this is done by the singlet representation on all subspaces, $(p_{\A}, q_{\A})=(p_{\B}, q_{\B})=(p_{\A\B}, q_{\A\B})=(0, 0)$, if it appears on the corresponding Hilbert spaces. According to Eq.~\eqref{eq:hspacedecomposition} this happens exactly when $N$ is divisible by 3. In the other cases, i.e., when   $\text{mod}(N, 3) \ne 0$, the ground state is labeled by small values of the $p$ and $q$ quantum numbers that do not scale with $N$: 
\begin{center}
  \begin{tabular}{cccc}
    $\text{mod}(N, 3)$& $(p_\A, q_\A)$& $(p_\B, q_\B)$& $(p_{\A\B}, q_{\A\B})$\\
    \hline
    0& $(0, 0)$& $(0, 0)$& $(0, 0)$\\
    1& $(1, 0)$& $(1, 0)$& $(0, 1)$\\
    2& $(0, 1)$& $(0, 1)$& $(1, 0)$
  \end{tabular}
  \captionof{table}{}\label{gstable}
\end{center}
The difference between the energy of these states and that of the singlet is also of order $O(1)$. The ground state of an antiferromagnetic Heisenberg model is a global singlet both on a complete graph connection layout and, according to Marshall's theorem \cite{marshall1955antiferromagnetism}, on a bipartite lattice with equal sized sublattices. The ground state we have here also falls in line with this behavior.
 \begin{figure*}
 \includegraphics[width=\textwidth]{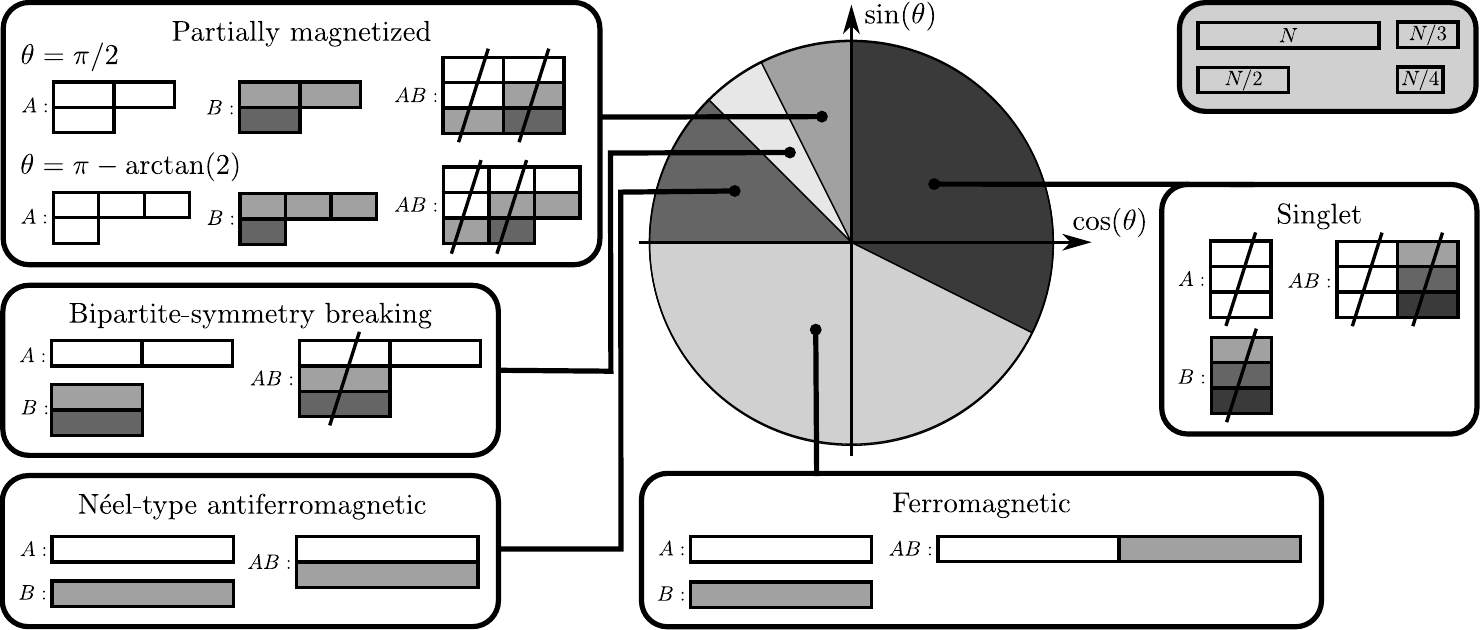}
    \caption{
      The ground-state phase diagram of the CBE Hamiltonian. Each ground-state subspace is labeled by three $\SU(3)$ Young diagrams corresponding to the two subsystems and the entire system. The different lengths of the rectangular blocks in the Young diagrams represent the number of boxes in the rows as shown in the legend in the top right corner. In order to display how the Littlewood-Richardson rules \cite{sagan2013symmetric} apply to the product diagram on the entire system (AB), we have colored the rows of the diagrams of the B subsystem and added a third line indicating $\SU(3)$-singlets in the system. With this third line, the number of boxes in each diagram is equal to the number of sites in the corresponding (sub)system. In order to recover the standard two-row $\SU(3)$ diagrams, one needs to remove all the columns with three boxes. This is indicated by these columns being crossed out. The diagram with all boxes crossed out corresponds to the label $(0, 0)$, i.e., the singlet representation. Starting from $\theta=0$ and going clockwise we have the singlet phase ($-\arctan((1+3/N)/(2+3/N))\le\theta\le\pi/2$), the ferromagnetic phase ($\pi\le\theta\le 2\pi-\arctan((1+3/N)/(2+3/N))$), the N\'eel-type antiferromagnetic phase ($3\pi/4\le\theta\le\pi$), the bipartite-symmetry breaking phase ($\pi-\arctan( 2(N+2)/(N+6)\le\theta\le 3\pi/4$), and the partially magnetized critical phase ($\pi/2\le\theta\le\pi-\arctan( 2(N+2)/(N+6)$). The ground state in the partially magnetized phase shifts many times with the value of $\theta$, in the diagram we only displayed the ground states at the two ends of the region (in the thermodynamic limit).}
   \label{fig:phdiag}
 \end{figure*}
\subsection{Third quarter $(\pi<\theta<3\pi/2)$}
The other region in the parameter space with no competition between the interactions across and within the subsystems is the third quarter, i.e., $0<\theta<\pi/2$. Here, we need to maximize the eigenvalues of all the Casimir operators. Consider the decomposition of the  Hilbert space of the entire system into $\SU(3)$ irreps; then, without regarding the restrictions coming from fixing the irreps on the A and B subsytems, the irrep that maximizes the eigenvalue of $C_{\A\B}$ is $(p_{\A\B}, q_{\A\B})=(2N, 0)$. The irreps that maximize the eigenvalues of $C_\A$ and $C_\B$ are $(p_{\A}, q_{\A})=(p_{\B}, q_{\B})=(N, 0)$. One can see from Eq.~\eqref{eq:productdecomposition} that $(2N, 0)$ is part of the irrep decomposition of $(N, 0)\otimes(N, 0)$. Therefore, the ground states of the CBE Hamiltonian in this quarter of the parameter space belong to the subspace labeled by the triple $( \mu_A,  \mu_B, \mu_{AB})=((N, 0),   (N, 0), (2N, 0))$. The Schur-Weyl duality \cite{weyl1939classical} gives us a straightforward interpretation of these numbers: the ground-state subspace is the symmetric part of the Hilbert space, spanned by vectors that are invariant to all permutations of sites. Since in this quarter both types of interactions are ferromagnetic, we expect the ground state to be similar to the ferromagnetic ground state of $SU(2)$ Heisenberg models. This matches both the interpretation from the Schur-Weyl duality and the maximal eigenvalues of the Casimir operators.

\subsection{Fourth quarter $(3\pi/2<\theta<2\pi)$} 
In this region of the parameter space, the two interactions in the CBE Hamiltonian \eqref{eq:HBLBQB} are competing with each other. The eigenvalues of the Casimir operators of the A and B subsystems needs to be minimized, while the Casimir for the entire Hilbert space needs to be maximized. According to Eq.~\eqref{eq:maxirrep} this latter means that  $\mu_{\A\B}^{\text{opt}}=(p_\A+p_\B,q_\A+q_\B)$. Using the new variables defined in Eq.~\eqref{eq:variabletransform} for the energy of the CBE Hamiltonian, we obtain
\begin{multline}
  \label{eq:3rdquartere}
  E=\left( \cos(\theta)+\sin(\theta) \right)\left(  c_{qu}(v_\A, x_\A)+c_{qu}(v_\B, x_\B) \right)+\\\sin(\theta)N^2v_\A v_\B\left(\frac{1}{3}+x_\A x_\B  \right).  
\end{multline}
From this one can see that the solution simplifies when the coefficients of both terms are negative, that is, in the region $3\pi/2<\theta<7\pi/4$. Here, the absolute value of both terms needs to be maximized, and the irreps that maximizes both is labeled by $x_\A=x_\B=1$ and $v_\A=v_\B=1$. In other words, the previously discussed ground state of the quarter $\pi<\theta<3\pi/2$ extends into this region.

This brings up two other questions: Could this symmetric ground state extend any further, and is it possible that the singlet ground state of the first quarter extends in a similar fashion into this parameter region? This last case could be feasible for values of $\theta$ for which the interaction  $C_\A+C_\B$ dominates the term $C_{\A\B}$. Since the energy of the singlet is $0$, it can be the ground state only when the energies of all other irrep combinations are positive. The inequality $E(v_\A, x_\A, v_\B, x_\B)\ge 0$ yields a condition for $\theta$ that has to apply to all possible values of $v_\A,x_\A,v_\B$ and $x_\B$:
\begin{equation}
  \label{3rdquarterineq}
  -\text{ctg}(\theta)\ge 1 + \frac{2+6 x_\A x_\B}{\frac{v_\A}{v_\B}(1+3x_\A^2)+\frac{v_\B}{v_\A}(1+3x_\B^2)}.
\end{equation}

In order to extract the critical value of parameter $\theta$, two observations should be made:
 First, when $v_\A=v_\B$, and $x_\A=x_\B$, the right-hand-side of  \eqref{3rdquarterineq} is equal to $2$; and second, by utilizing $x a+b/x \ge 2\sqrt{b a}$ and the inequality between the arithmetic and geometric means, one obtains that the right hand side of \eqref{3rdquarterineq} always has to be less or equal to $2$.
Thus, the singlet subspace is the ground-state subspace in this region if and only  if   $  -\text{ctg}(\theta)\ge 2 $, which means that it extends  from $\theta=2\pi$ until $\theta = 2\pi-\arctan(1/2)$.

Next, we check whether the symmetric ground state extends any further. The inequality $E(1, 1, 1, 1)\le E(v_\A, x_\A, v_\B, x_\B)$ provides the following condition for $\theta$:
\begin{equation}
  \label{4thquarterineq}
  -\text{ctg}(\theta) \le 1 + \frac{2v_{\A} v_{\B} (1 + 3 x_{\A} x_{\B})-8}{v^2_{\A}(1+3x_\A^2)+v^2_{\B}(1+3x_\B^2)-8}.
\end{equation}

We follow a reasoning analogous to the one after Eq.~\eqref{3rdquarterineq}. Using the relation between the geometric and arithmetic means, it is easy to see that $2$ is a strict lower bound of the right hand side of Eq.~\eqref{4thquarterineq}. Moreover, the right hand side reaches this lower bound iff $x_{\A}=x_{\B}$ and $v_{\A}=v_{\B}$. We conclude that the symmetric ground state extends until $\theta=2\pi-\arctan(1/2)$ and therefore, there is a direct transition between the symmetric and the singlet ground states at this parameter value.

\subsection{Second quarter $(\pi/2<\theta<\pi)$} 
In the remaining quarter of the parameter space, $\pi/2<\theta<\pi$, the interactions within and across the subsystems are again competing. This time the coefficient of $C_{\A\B}$ in the CBE Hamiltonian \eqref{eq:HBLBQB} is positive, therefore, we use the irrep corresponding to the minimal eigenvalue of $C_{\A\B}$, $\mu_{\A\B}^{\text{opt}}=\check{\mu}_{\A\B}$, and substitute Eq.~\eqref{eq:cabmin} into the energy,
\begin{multline}
  \label{eq:energyinterestingquarter}
  E=\left( \cos(\theta)+\sin(\theta) \right)\left( c_{qu}(v_\A, x_\A)+c_{qu}(v_\B, x_\B) \right)-\\\sin(\theta)\frac{N^2}{6}v_1v_2\left(1+3x_1+3x_2-3x_1x_2 \right).
\end{multline}
It is clear that when the coefficients of both terms are negative, that is, when $3\pi/4<\theta<\pi$, the irreps on the A and B subspaces which minimize this expression are labeled by $v_\A=v_B=x_\A=x_\B=1$. However, this ground-state subspace is not an extension of that of the quarter $\pi<\theta<3\pi/2$, even though the $x$ and $v$ paremeters are identical: In the said case, the prefactor  of $C_{\A\B}$ is negative and the ground state corresponds to $\mu_{\A}=\mu_\B=(N,0)$ and $\mu_{\A\B}=(2N,0)$. Contrarily, in the present case, we have to choose the $\SU(3)$ irrep in the product $(N, 0)\otimes (N, 0)$ that corresponds to the minimal eigenvalue of the $C_{\A\B}$ Casimir operator, which according to Eq.~\eqref{eq:cabmincases} is $\mu_{\A\B}=(0, N)$. This ground state is similar to a N\'eel-type antiferromagnetic order in the sense that the two sublattices of a bipartite lattice are ferromagnetically aligned, but the value value of the quadratic Casimir operator on the entire lattice, is minimized.

In the remaining part of the domain of $\theta$, i.e., $\pi/2<\theta<3\pi/4$, finding the ground state becomes somewhat more complicated. Unlike the previous cases, we cannot immediately tell the value of the $v_A$ and $v_B$ variables in the ground state. Instead, we have to find the minima of a polynomial of four variables on the convex set describing the domain of these variables. Using a scaling argument, we can reduce the number of variables to three. First, we remark that for a suitably large value of $N$ the ground state energy of the CBE Hamiltonian is guaranteed to be negative in the parameter region we are currently investigating. Indeed, in the case of $\text{mod}(N, 3)=0$ there exists at least one  combination of irreps for which the energy is negative, we select a pair of conjugate representations, $(p, q)$ and $(q, p)$ on the $A$ and $B$ subspaces. The product of these contains the $\SU(3)$ singlet $(0, 0)$, thus the contribution of the term proportional to $C_{\A\B}$ to the energy is $0$ \footnote{In the cases when $\text{mod}(N, 3)\neq 0$, we can always choose a pair of irreps that are almost conjugates of each other and for which the term proportional to $C_{\A}+C_{\B}$ has an energy contribution of $O(N^2)$. In the irreducible decomposition of the direct product of these two irreps, the irrep with the lowest $C_{\A\B}$ eigenvalue is either $(1, 1)$ or $(1, 0)$, both of which have an energy contribution of $O(1)$. Thus, we can again conclude that the ground-state energy has to be negative.}. Second, we use the fact that the ground-state energy is negative to get rid of one variable in the optimization problem. Assume that in the ground state $v_\A\le v_\B$. Since $c_{qu}(v_\A, x_\A)$ contains only terms proportional to $N^2$, the ground-state energy given by Eq.~\eqref{eq:energyinterestingquarter} scales quadratically when we scale both $v_\A$ and $v_\B$ by the same constant, hence,
\begin{multline}
 E(\frac{v_\A}{v_\B},x_\A,1,x_\B)=\frac{1}{v_\B^2}E(v_\A, x_\A, v_B, x_B)\le\\  E(v_\A, x_\A, v_B, x_B),
\end{multline}
where the last inequality holds as $E$ is negative and $v^2_B \le 1 $.
It follows that when searching for the ground state we can set $v_B$ to $1$. In the following we determine the minimum of the polynomial $E(v_\A, x_\A, 1, x_\B)$ inside the domain of the remaining three variables. This minimum has different qualities depending on the value of $\theta$.

In the region $\pi/2<\theta<\pi-\arctan{2}$ the minimum inside the domain of the variables is a local minimum of the polynomial $E(v_\A, x_\A, 1, x_\B)$. At this local minimum $v_\A=v_\B=1$, and $x_\A=x_\B=x(\theta)$, a smooth function of $\theta$. Up to this stage of the calculation, we could handle $x_\A$ and $x_\B$ as continuous variables. Yet, when extracting the discrete $(p, q)$ values labeling the ground state, we need to take into account that in the case of $v_\A=v_\B=1$ they can only take the values $(\text{mod}(N,2)+2i)/N$, with $i$ being an integer between $0$ and $N/2$; That is, among the two proper values neighbouring $x(\theta)$, the ground state is the one with the lower energy. Since the energy Eq.~\eqref{eq:energyinterestingquarter} as a function of $x_\A=x_\B=x$ is a parabola, we can simply round $x(\theta)$ to its closest integer value. After doing this and using Eq.~\eqref{eq:cabmin} to determine the corresponding irrep, $(p_{\A\B}, q_{\A\B})$, on the entire Hilbert space, we arrive at the irreps labeling the ground state \footnote{For simplicity, we gave here the result for the case when $N$ is even. For the case when $N$ is odd, the ground-state labels are given by almost the same equations, only the floor function has to be used instead of rounding the terms to the nearest integer in Eq.~\eqref{eq:contgs}, furthermore one has to add the constant $1$ to $p_\A$ and $p_\B$, $-1/2$ to $q_\A$ and $q_\B$, and $3/2$ to $q_{\A\B}$.}:
  \begin{align}
    \label{eq:contgs}
    \begin{split}
      p_\A=&p_\B=2\Bigg{\lceil} \frac{1}{2}\left(  \frac{N}{3+2\text{ctg}(\theta)} \right)\Bigg{\rfloor},\\
      q_\A=&q_\B=\frac{N}{2}-\Bigg{\lceil} \frac{1}{2}\left(  \frac{N}{3+2\text{ctg}(\theta)}\right)\Bigg{\rfloor},\\
      p_{\A\B}=&0,\\
      q_{\A\B}=&3\Bigg{\lceil} \frac{1}{2}\left(  \frac{N}{3+2\text{ctg}(\theta)} \right)\Bigg{\rfloor}-\frac{N}{2},
    \end{split}
  \end{align}
 where $\lceil x \rfloor$ denotes closest integer value of $x$.
  
 In the region $\pi-\arctan{2}<\theta<3\pi/4$, the polynomial $E(v_\A, x_\A, 1, x_\B)$ has no local minimum inside the domain of its variables, therefore, the minimum has to be on the border of the domain. In fact there are two minima occupying two different extremal points of the domain, they are located at $(v_\A, x_\A, v_\B, x_\B)=(1, 1, 1, 0)$ and $(v_\A, x_\A, v_\B, x_\B)=(1, 0, 1, 1)$. The most peculiar quality of the ground states associated with these minima is that unlike all previously discussed ground states, they break the bipartite symmetry of the CBE Hamiltonian. This also explains why these ground states come as a pair, when the $A$ and $B$ subsystems are swapped the two minima are transformed into each other. After taking into account the discrete nature of our variables and rounding the location of the minima appropriately, then extracting $p_{\A\B}$ and $q_{\A\B}$ from Eq.~\eqref{eq:cabmin}, we arrive at the two sets of $\SU(3)$ irreps labeling the ground state. The first one is $(p_\A, q_\A)=(N, 0)$, $(p_\B, q_\B)=(\text{mod}(N, 2),\lfloor N/2 \rfloor)$, $(p_{\A\B}, q_{\A\B})=(\lfloor N/2 \rfloor,\text{mod}(N, 2))$, and the second one is obtained from the former by swapping the A and B subsystems.

\subsection{Special parameter values}

For generic values of the parameter $\theta$, the ground-state subspace of the CBE Hamiltonian \eqref{eq:HBLBQB} belongs to a fixed set of quantum numbers, i.e., irrep labels $(\mu_\A,\mu_\B,\mu_{\A\B})$. However, at the borders of the different phases, the ground-state subspace becomes more degenerate incorporating states with different irrep labels, or in other words, multiple sets of quantum numbers become degenerate in energy.  For example, at the borders of the phases at least two sets of labels correspond to the ground-state energy, but further degeneracies are also possible depending on the form the energy takes at the given parameter. If this happens we have to keep in mind that when determining the ground-state energies, we have neglected the parts of the Casimir operators eigenvalue, Eq.~\eqref{eq:Casimir}, that are only linear in $N$. So far this has been acceptable because we were only interested in the thermodynamic limit, and the other terms scale with $O(N^2)$. However, at the values of $\theta$ where the quantum numbers describing the ground-state subspace become degenerate, there is a possibility that the linear terms break the degeneracy.  We should also note that at the parameter values $\theta=0, \pi/2, \pi, 3 \pi/2$ either the $\mu_{AB}$ or the $\mu_A$ and $\mu_B$ ceases being a relevant quantum number which could also lead to degeneracies; two of these values ($\theta=\pi/2, \pi$) are also phase boundaries, but the other two should be considered separately. In this subsection, we check each of these special parameter values.

Let us start with the two special points that are not at a phase boundary, i.e., $\theta=0$ and $\theta= 3 \pi/2$. At $\theta=0$, the irreps $\mu_\A$ and $\mu_\B$ labeling the ground state are the same as those inside the singlet phase, listed in Table~\ref{gstable}. However, since at this point the CBE Hamiltonian is governed solely by the interaction within the A and B subsystems, $\mu_{\A\B}$ stops being a relevant quantum number and the ground-state subspace extends to the entire $\mu_\A\otimes\mu_\B$ subspace. In practice, this means that there is no additional degeneracy when $\text{mod}(N,3)=0$, but in the other two cases the ground-state subspace is slightly enlarged. The situation at $\theta=3\pi/2$ is in some sense the dual to the previous case, as the CBE Hamiltonian  takes the form $H=-C_{\A\B}$, and the only relevant label is $\mu_{\A\B}$. However, since the irrep $\mu_{\A\B}=(2N,0)$ is compatible only with the irreps $\mu_\A=\mu_\B=(N,0)$ on the subsystems, there is no additional degeneracy of the ground state.

The parameter value  $\theta=\pi/2$ is at the boundary of the singlet and the partially magnetized  phases. Here, the CBE Hamiltonian takes the form $H=C_{\A\B}$, thus, $\mu_\A$ and $\mu_\B$ are not relevant labels of the energy eigenstates. The irrep $\mu_{\A\B}$ corresponding to the ground state is the one appearing in Table \ref{gstable}, and the ground state is extended to the entire $\mu_{\A\B}$ subspace. Compared to the  case at $\theta=0$, the degeneracy here is a lot more extensive.
At the boundary point of the ferromagnetic and N\'eel-type antoferromagnetic phase, $\theta=\pi$, the CBE Hamiltonian takes the form $H=-(C_\A+C_\B)$. Here, since only $\mu_\A$ and $\mu_\B$ are relevant labels, the ground-state subspace is enlarged to the entire $(N,0)\otimes(N,0)$ subspace.

At the point $\theta=2\pi-\arctan(1/2)$, where the singlet and the ferromagnetic phases meet, the expression of the energy in the fourth quarter, shown in Eq.~\eqref{eq:3rdquartere} takes the form:
\begin{equation}
  E=\frac{\sqrt{5}}{30}N^2\left[ (v_\A-v_\B)^2+3(v_\A x_\A-v_\B x_{\B})^2 \right ].
\end{equation} 
At this point the ground-state subspace encompasses all irreducible subspaces for which the two subsystems are symmetric to exchange and the energy contribution of $C_{\A\B}$ is maximized. In other words $(p_\A, q_\A)=(p_\B, q_\B)=(p,q)$ and $(p_{\A\B}, q_{\A\B})=(2p, 2q)$. In this situation however, we must take into account the previously omitted parts of the energy that are linear in $N$, since these might break the degeneracy. By checking the energy contributions of these linear terms one can make two important conclusions. First, the value of the parameter $\theta$ where the shift between the two types of ground state occurs receives a correction for finite values of $N$, $\theta=2\pi-\arctan((1+3/N)/(2+3/N))$. Second, the degeneracy is broken, and the ground-state subspace consists only of the two types of ground states neighboring the critical point: the singlet subspace and the symmetric subspace.

The N\'eel-type antiferromagnetic and the bipartite symmetry breaking phases border at $\theta=3\pi/4$, here the CBE Hamiltonian is proportional to $C_{\A\B}-C_{\A}-C_{\B}$. The subspace where the ground-state energy, Eq.~\eqref{eq:energyinterestingquarter}, is minimal is larger than the span of the ground-state subspaces of the two adjecent phases. It encompasses all subspaces with labels of the form $(p_\A, q_\A)=(N,0)$, $(p_\B, q_\B)=(Nx, N/2(1-x))$ $(p_{\A\B}, q_{\A\B})=(N/2(1-x), Nx)$ with $x\in[0,1]$, and those one gets form the former set by swapping the $\A$ and $\B$ subsystems. The energy contribution of the $O(N)$ parts of the Casimir operators is constant in the entire ground-state subspace, therefore this degeneracy remains.

The border of the bipartite symmetry breaking and the partially magnetized phases is at $\theta=\pi-\arctan{2}$. At this point, according to the part of the energy that scales quadratically with $N$, the ground-state subspace is the span of a number of irreducible subspaces which break the bipartite symmetry. The labels for these take the form $(p_\A,q_\A)=(Nx, N(1-x)/2)$, $(p_\B,q_\B)=(N(1-x),Nx/2)$ and $(p_{\A\B}, q_{\A\B})=(N|1/2-x|, N/2(1/2-|1/2-x|))$, with $x\in[0,1]$.  However, the energy contribution of the linear terms breaks this degeneracy and, away from the thermodynamic limit, adjusts the critical parameter value where the ground-state phases change by a correction of magnitude $O(1/N)$. The new value is $\theta=\pi-\arctan( 2(N+2)/(N+6) )$ and the ground-state subspace is the span of the ground-state subspaces of the two adjacent phases.

\subsection{Energy gaps} 

From a many-body point of view it is important to know whether the different quantum phases of our model are gapped or gapless in the thermodynamic limit. As a consequence of the infinite range interaction the eigenvalues of the CBE Hamiltonian \eqref{eq:HBLBQB} are not extensive quantities. In order to make the energy extensive and meaningfully define a gap, we should normalize the Hamiltonian by a factor of $1/N$, which is a usual procedure in models on complete graphs \cite{friedli_velenik_2017}.   

\begin{figure}
  \centering
    \includegraphics[width=0.4\textwidth]{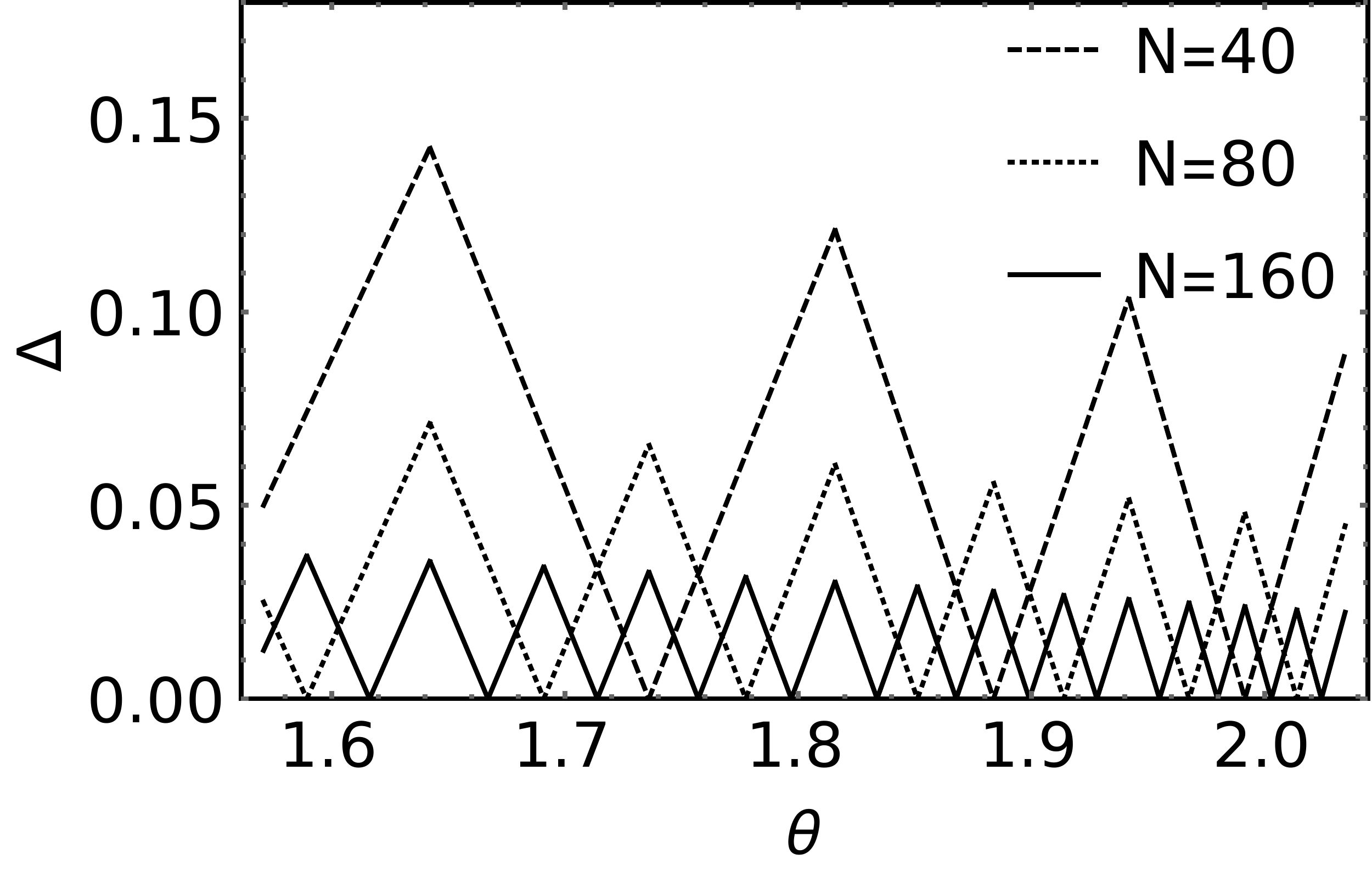}
   \caption{The normalized energy gap $\Delta$ in the parameter region $\pi/2\le\theta\le\pi-\text{arctan}(2)$ for different system sizes.}
   \label{fig:gap}
 \end{figure}

Let us now investigate the energy gaps taking into account the normalization factor. In the singlet phase, corresponding to the region $-\arctan((1+3/N)/(2+3/N))\le\theta\le\pi/2$, the Casimir eigenvalues corresponding to both the ground-state subspace and the states with the second lowest energy are constant in $N$ (apart from the $mod \, 3$ oscillations), therefore, the normalized CBE Hamiltonian in this phase is gapless. The three different phases in the parameter region $\pi-\arctan( 2(N+2)/(N+6) )\le\theta\le 2\pi-\arctan((1+3/N)/(2+3/N))$ have the unifying feature that the Casimir eigenvalues of the ground-state subspace and the second lowest energy states (which we can obtain from the ground state by a small constant modification of the appropriate quantum numbers) are of order $O(N^2)$, and their difference is of order $O(N)$. Taking the normalization into account, we obtain that these phases are gapped. Finally, in the parameter region $\pi/2\le\theta\le\pi-\arctan(2(N+2)/(N+6))$, the quantum numbers describing the ground state change many times. The behavior of the gap in this phase is shown in Fig.~\ref{fig:gap}. According to Eq.~\eqref{eq:contgs}, there is a ground state level crossing at each $\theta$ where the number $N/(6+4\text{ctg}(\theta))$ is half-integer. The state with the second lowest energy is always given by rounding the number $N/(6+4\text{ctg}(\theta))$  in Eq.~\eqref{eq:contgs} to the next closest integer. The density of these level crossings increases linearly with $N$. Additionally, the local maximums of the gap between the level crossings are enveloped by a smooth function which gives us an upper bound for the value of the gap $\Delta$:
\begin{equation}
  \label{eq:gapenvelope}
\Delta \le\frac{1}{N} \left(4\cos(\theta)+6\sin(\theta)\right).
\end{equation}
Therefore, the continuous phase is gapless in the thermodynamic limit.

  \section{Summary and  Outlook}
\label{sec:summary}

  Studying spin systems on complete graphs has a long history in many-body physics. Such models have been considered in the past  mainly as infinite-dimensional mean-field versions of their finite-dimensional lattice counter-parts, examples include  the Curie-Weiss-model \cite{friedli_velenik_2017} and the Sherrington-Kirkpatrick version of spin-glass models \cite{sherrington1975solvable}. With the advent of cold atomic systems, long-range interactions, including complete-graph interactions could also be realized in the lab \cite{beverland2016realizing}. In this paper, we have considered a slight modification of this approach by studying a quantum spin model on a bipartite complete graph, which could be regarded as a mean-field approach that captures also the effects that stems from the bipartition of a lattice. Moreover, such a bipartite model might also be realized using experimental techniques with ultracold atoms and cavity electrodynamics. 
  
  We have identified five quantum phases of this model, as shown in  Fig.~\ref{fig:phdiag}. There are two gapless phases, the antiferromagnetic singlet phase and the partially magnetized critical phase; and three gapped phases, the ferromagnetic phase, the N\'eel-type of antiferromagnetic phase with ferromagnetically aligned subsystems, and a bipartite-symmetry-breaking phase. Concerning this last phase, it is interesting to note, that already such a simple bipartite long-range model provides a phase that is absent  in the literature on short-ranged bipartite models. In this phase, although the two subsystems transform under the same representation of $\SU(3)$, the ground state of this phase restricted to the subsystems belongs to different representations.
  
  There are a number of ways how one can extend the present study. A straightforward modification would be to consider subsystems with different sizes, in particular, one could study the limiting case of a central spin (or spin-star) model, where one subsystem is simply a single spin-$1$ particle. The topology of the couplings could also be changed more drastically, for example by extending the bipartite system discussed here into a multipartite mean-field model by considering $k$ subsystems with collective spin-spin interactions within and across the subsystems. Furthermore, one could also relax the complete connectivity, and study similar models with decaying long-range interactions. Another interesting generalization that does not involve the spatial redistribution of the couplings would be to study less symmetric interactions; a natural candidate would be reducing the $\SU(3)$ symmetry to $\SU(2)$. In the case of spin-1 this is described by the bilinear-biquadratic interaction, which has already been studied on complete graphs \cite{jakab2018bilinear}. Symmetric collective spin states have been studied, due to their experimental feasibility, also from a quantum metrology point of view \cite{urizar2013macroscopic, apellaniz2018precision}, it would be interesting to study also bipartite models especially in the light of the experiment reported in \cite{lange2018entanglement}.
  A further direction would be to investigate not only static properties, but time-evolutions, e.g., different quench protocols. Such quench studies would also be of great interest if one would be able to experimentally realize such collective models, as discussed earlier, and  then observe the quench dynamics in the lab.

  \section*{Acknowledgements}
  
 We thank K. Penc, A. Solymos, and particularly G. Szirmai for helpful discussions. We acknowledge support from the Hungarian National Research, Development and Innovation Office (NKFIH) through Grants No. K124351, K124152, K124176 KH129601, and the Hungarian Quantum Technology National Excellence Program (Project No. 2017-1.2.1-NKP-2017-00001); and ZZ was also partially funded by the J\'anos Bolyai and the Bolyai+ Scholarships.
  
  \appendix
  \section*{Appendix}
  \section{The optimal eigenvalues of $C_{\A\B}$}
  \label{appendix:eigenvalues}
  Here, we identify the irreps $(\hat{p}_{\A\B}, \hat{q}_{\A\B})$ and $(\check{p}_{\A\B}, \check{q}_{\A\B})$ in the irreducible decomposition of $(p_\A, q_\A)\otimes(p_\B, q_\B)$ that  respectively maximize and minimize the eigenvalue of the Casimir operator $C_{\A\B}$ specified in Eq.~\eqref{eq:Casimir}. 
  
  We begin with determining the  irrep  $(\hat{p}_{\A\B}, \hat{q}_{\A\B})$  that maximizes the eigenvalue of $C_{\A\\B}$. The number of boxes in the Young diagram corresponding to $(p, q)$ is $n=2q+p$. We let $p$ and $q$ in Eq.~\eqref{eq:Casimir} vary with the restriction of keeping $n$ constant,
  this way, $c(p, q)|_{n=\text{const}}$ is monotonically increasing with the length of the first row of the diagram, $\nu=p+q$. 
  According to Eq.~\eqref{eq:productdecomposition}, the diagram in the irrep decomposition of $(p_\A, q_\A)\otimes(p_\B, q_\B)$ with the longest first row is $(p_\A+p_\B, q_\A+q_\B)$, however, the diagrams in the right hand side of Eq.~\eqref{eq:productdecomposition} have varying numbers of boxes. We now prove that regardless, this is the diagram that we are looking for. The number of boxes in a particular diagram on the right hand side of Eq.~\eqref{eq:productdecomposition} is
  \begin{equation}
    n = p_\A+p_\B+2(q_\A+q_\B)-3(i+l). 
  \end{equation} 
  Once again, we note that this number can only change in multiples of three. This property can be explained more deeply by the Schur-Weyl duality \cite{weyl1939classical} and the connection between the irreps of \SU(3) and \U(3) \cite{iachello}. Now let us consider those diagrams on the right hand side of Eq.~\eqref{eq:productdecomposition} for which $v=i+l$, and by extension the number of boxes takes a fixed value. For these, the length of the first row is
  \begin{equation}
    \nu|_{v=\text{const}}(i, k)=p_\A+q_\A+p_\B+q_\B-v-i-k.
  \end{equation}
  The diagram corresponding to $i=k=0$ is $(p_\A+p_\B+v, q_\A+q_\B-2v)$; note that depending on the value of $v$, this diagram does not necessarily appear in Eq.~\eqref{eq:productdecomposition}. We denote by $(\hat{p}_v, \hat{q}_v)$ the diagram in the right hand side of Eq.~\eqref{eq:productdecomposition} with a fixed value of $v$, which maximizes $c(p, q)$. Since $i>0$ and $k>0$ we have \begin{equation}
    c(p_\A+p_\B+v, q_\A+q_\B-2v)>c(\hat{p}_v, \hat{q}_v).
  \end{equation}
  In order to prove the initial statement, we need to show that
  \begin{equation}
    c(p_\A+ p_\B, q_\A+q_\B)\ge c(p_\A+p_\B+v, q_\A+q_\B-2v),
  \end{equation}
  which, after substituting to Eq.~\eqref{eq:Casimir}, reduces to
  \begin{equation}
    \label{eq:vineq}
    1+q_\A+q_\B\ge v .
  \end{equation}
  Using upper bounds of the indices $i$ and $l$ in Eq.~\eqref{eq:productdecomposition}, one gets an upper bound for $v$,
  \begin{equation}
    v=i+l\le \text{min}(p_\B, q_\A)+\text{min}(q_\A+k-i, q_\B)\le q_\A+q_\B. 
  \end{equation}
  Therefore, the inequality \eqref{eq:vineq} is satisfied for every irrep on the right hand side of Eq.~\eqref{eq:productdecomposition}. We conclude that $(\hat{p}_{\A\B}, \hat{q}_{\A\B})=(p_\A+p_\B, q_\A+q_\B)$.\\

  Now, we identify the irrep $(\check{p}_{\A\B}, \check{q}_{\A\B})$ in the irreducible decomposition of $(p_\A, q_\A)\otimes(p_\B, q_\B)$ that minimizes the eigenvalue of $C_{\A\\B}$. If we regard the indices $i, k$ and $l$ in Eq.~\eqref{eq:productdecomposition} as continuous variables, and substitute the $p(i, k, l)$ and $q(i, k, l)$ values of the irreps in the decomposition into Eq.~\eqref{eq:Casimir}, then it is straightforward to see that the resulting polynomial of the variables $i,k,l$ has no local minima inside the region specified by the bounds in Eq.~\eqref{eq:productdecomposition}. This means that the irrep we are looking for is on the border of the region. According to Eq.~\eqref{eq:productdecomposition}, the change in the eigenvalue of the Casimir operator $C_{\A\B}$ when we increase $l$ by 1 while keeping $i$ and $k$ constant is always negative, i.e., 
  \begin{equation}
    c(i, k, l+1)-c(i, k, l)=-3q-1<0.
  \end{equation}
  Thus, we need to search on the part of the border of the $i$, $k$, $l$ parameter region where $l$ is maximal, i.e., $l=l_1$. In a similar fashion, it is easy to show that,
  \begin{gather}
    c(i, k+1, l_1)-c(i, k, l_1)<0,\quad \text{and},\\  c(i+1, k_1, l_1)-c(i, k_1, l_1)<0.
  \end{gather}
  Consequently, for the irrep that minimizes the eigenvalue of $C_{\A\B}$ all three indices take their maximal values, $i=i_1$, $k=k_1$ and $l=l_1$. In order to express the $\check{p}_{\A\B}$ and $\check{q}_{\A\B}$ values corresponding to these indices using $(p_\A, q_\A)$ and $(p_\B, q_\B)$, we introduce $X=p_\A-q_\B$ and $Y=p_\B-q_\A$. With these,
  \begin{multline}
    (\check{p}_{\A\B}, \check{q}_{\A\B})=
    \begin{cases}
      (X-Y, Y)\quad\text{if}\  Y>0\ \text{and}\  X>Y \\
      (Y-X, X)\quad\text{if}\  X>0\ \text{and}\  X\le Y\\
      (Y, -X)\quad\text{if}\  X\le 0\ \text{and}\  Y>0\\
      (X, -Y)\quad\text{if}\  X>0\ \text{and}\  Y\le 0\\
      (-X, X-Y)\quad\text{if}\  X\le 0\ \text{and}\  X>Y\\
      (-Y, Y-X)\quad\text{if}\  Y\le 0\ \text{and}\  X\le Y.
    \end{cases}
  \end{multline}\\[5mm]
  \bibliography{blbq}

\end{document}